\documentclass{elsart}

\usepackage{psfig}

\begin{document}
  
  \begin{frontmatter}
  
    \title {Complete Two Loop Electroweak Contributions to the Muon
    Lifetime in the Standard Model}

    \author{M. Awramik$^1$} and \author{M. Czakon$^2$}

    \address{$^1$ Institute of Nuclear Physics, Radzikowskiego 152,
      PL-31342 Cracow, Poland}

    \address{$^2$ Department of Field Theory and Particle Physics,
      Institute of Physics, University of Silesia, Uniwersytecka 4,
      PL-40007 Katowice, Poland}
  
    \begin{abstract}

      An independent result for the two loop fermionic contributions
      to the muon lifetime in the  Standard Model is
      obtained. Deviations are found with respect to
      \cite{Freitas:2000gg}, which result in a  shift of the $W$ boson
      mass by approximately $-1.3$~MeV over the range of Higgs boson
      masses from 100~GeV to 1~TeV. Supplied with the bosonic
      contributions from
      \cite{Awramik:2002wn,Onishchenko:2002ve,Awramik:2002vu}, this
      shift, due to the complete electroweak contributions, varies
      from $-2.4$~MeV to $-0.6$~MeV. Additionally, a new test of the
      matching procedure defining the Fermi constant is presented,
      which uses fermion masses as infrared regulators.
    
    \end{abstract}

  \end{frontmatter}
  
  \section{Introduction}

  The muon lifetime is one of the key observables of today's particle
  physics. Not only is it measured very precisely, since the current
  experimental error is 18 ppm \cite{Hagiwara:fs}, but can be
  described to competing accuracy within the Standard Model, giving
  rise to a strong correlation between the masses of the heavy gauge
  bosons. As a low energy process, the decay is expected to be
  governed by an effective interaction involving only the electron,
  muon and their respective neutrinos. The dynamics of the system
  should be corrected mostly by QED, whereas the electroweak
  interactions determine solely the size of the coupling constant.

  The history of the calculation of the electroweak corrections, in
  which we are interested here, is rather long. It started in the
  early eighties with the one loop contributions
  \cite{Sirlin:1980nh}. Subsequently, leading terms in the top quark
  ${\mathcal O}(\alpha^2 m_t^4)$ \cite{vanderBij:1986hy} and Higgs
  boson ${\mathcal O}(\alpha^2 M_H^2)$ \cite{vanderBij:1983bw} masses
  were derived at the two loop level. In the meantime, mixed
  electroweak and QCD corrections became available at order ${\mathcal
  O}(\alpha \alpha_s)$ \cite{Djouadi:gn} and ${\mathcal O}(\alpha
  \alpha_s^2)$ \cite{Avdeev:db}. Recently, three loop leading terms in
  the top quark mass ${\mathcal O}(\alpha^3 m_t^6)$ and ${\mathcal
  O}(\alpha^2\alpha_sm_t^4)$ have also been calculated
  \cite{Faisst:2003px}. As far as the pure two loop electroweak
  corrections are concerned, after it turned out that the subleading
  terms in the top quark mass expansion are comparable with the
  leading ones \cite{Degrassi:1996mg}, complete fermionic and the
  Higgs boson mass dependence of the bosonic contributions have been
  evaluated \cite{Freitas:2000gg}. The complete bosonic part has been
  done in \cite{Awramik:2002wn,Onishchenko:2002ve,Awramik:2002vu}. It
  is the purpose of the present paper to present the result of a new
  independent calculation of the fermionic contributions and, after
  inclusion of the bosonic corrections, also of the full electroweak
  corrections.

  This work is organized as follows. In the next Section, we discuss
  the matching procedure and fermion masses as infrared regulators,
  which avoid ambiguities of the definition of gamma matrices in box
  diagrams in noninteger dimensions. Then, we present the results for
  the fermionic and full contributions and compare them with previous
  calculations by specifying the differences in the $W$ boson mass
  prediction. Conclusions close the paper.

  \section{Matching}

  Due to a large number of very different mass scales, it is virtually
  impossible to evaluate the muon decay lifetime directly within the
  Standard Model, without recourse to some approximation method. An
  elegant and systematic approximation is provided by the approach
  based on {\it effective theories}. The idea is to agree on some
  cutoff scale, below which all degrees of freedom are treated
  exactly, whereas the heavier fields are ``integrated out'', which
  means that they generate effective interactions. It should not be
  surprising that one first discovers experimentally the effective
  theories, since the dependence on the heavier scales requires higher
  ``resolution'', {\it i.e.} higher energy. For precisely this reason,
  the effective theory governing muon decay, the Fermi Model, has been
  known much before the Standard Model. From this point of view, one
  should not consider that the Fermi Model is used in current
  calculations for {\it historical reasons}, but because it is the
  appropriate effective theory at this energy scale.

  The approximation is constructed as follows. The lagrangian is made
  only from the light fields, which are the six leptons, the five
  quarks, the photon and the gluon. At leading order in the inverse
  heavy scale, for which we take the $W$ boson mass $M_W$, a single
  effective operator is added, giving the lagrangian (in the
  so--called charge conserving form of the Fermi operator)
  \begin{eqnarray}
    \label{Leff}
	  {\mathcal L}_{\rm eff} &=& {\mathcal L}_{\rm kin}(\nu)+
	  {\mathcal L}_{\rm QED}(\alpha^0,m_l^0,m_q^0,l^0,q^0,A^0_\mu)
	  +{\mathcal L}_{\rm QCD}(\alpha_s^0,m_q^0,q^0,A^{a,0}_\mu) 
	  \\ \nonumber
	  &+& \frac{G_F}{\sqrt{2}}\;\;\overline{e^0}
	  \gamma^\alpha (1-\gamma_5) \mu^0 \times \overline{\nu_\mu}
	  \gamma_\alpha (1-\gamma_5) \nu_e,
  \end{eqnarray}
  where the superscript $0$ denotes bare quantities. The theory is
  finite after mass and coupling ($\alpha$ and $\alpha_s$)
  renormalization to all orders in $\alpha$ and $\alpha_s$, and
  leading order in the Fermi constant $G_F \sim 1/M_W^2$, which is why
  this parameter is not renormalized.

\begin{figure}
    \psfig{file=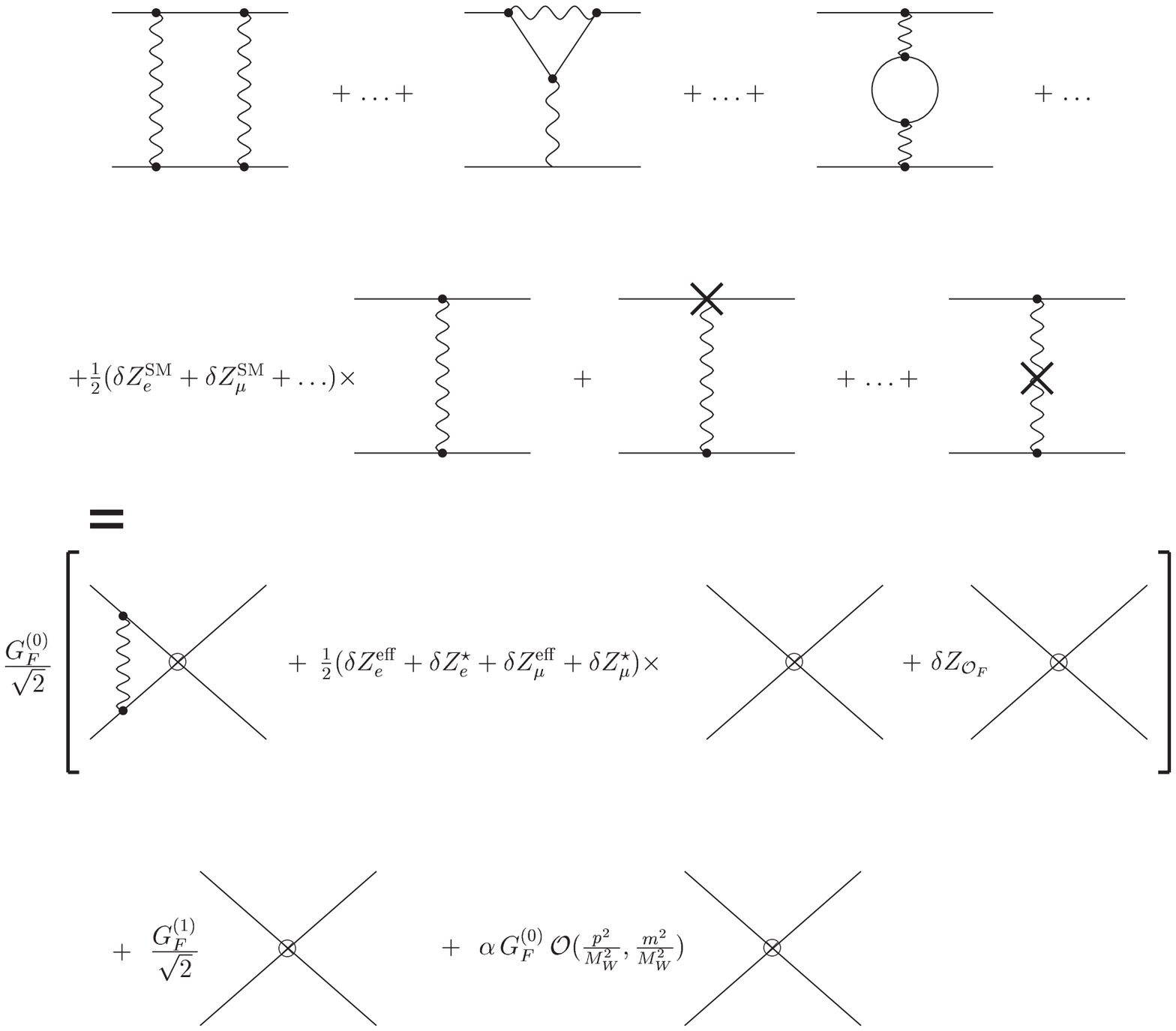,width=13cm}
    \caption{\label{matching1l} Matching equation at the one loop
      level. The wavy lines on the left hand side represent the three
      gauge bosons, $\gamma$, $W$ and $Z$, whereas on the right, only
      the photon. The rest of the notation is explained in the text.}
  \end{figure}

  The {\it matching procedure} in the present case consists in
  requiring that the amputated renormalized Green
  functions\footnote{This choice is somewhat arbitrary, since one
  might just as well use full Green functions, or Green functions
  which are one particle irreducible with respect to the light
  fields.} of the effective theory be equal to the amputated
  renormalized Green functions of the Standard Model up to terms of
  order ${\mathcal O}(1/M_W^4)$ and given order in $\alpha$ and
  $\alpha_s$
  \begin{equation}
    \label{matching}
    {\mathcal G}_{\rm SM} = {\mathcal G}_{\rm eff}
    +{\mathcal O}(1/M_W^4),
  \end{equation}
  which makes the muon decay amplitude the same in both models up to
  the specified order. The Fermi constant is then given as an
  expansion in $\alpha$ and $\alpha_s$
  \begin{equation}
    G_F=\sum_{i=0}^{\infty} G_F^{(i)}=\frac{\pi \alpha}{\sqrt{2} s_W^2
    M_W^2}(1+\Delta r),
  \end{equation}
  with $G_F^{(0)} = \pi \alpha/(\sqrt{2} s_W^2 M_W^2)$ being the Born
  level prediction. The quantity $\Delta r$ is customarily used to
  parametrize the higher order contributions. At the one loop level,
  the matching equation is schematically depicted in
  Fig.~\ref{matching1l}.
  We introduced there the decoupling coefficients
  \cite{Weinberg:1980wa}, $Z^*_{e,\mu}$, which are different from one
  in the $\overline{\rm MS}$ scheme for example, but can be neglected
  in the on-shell scheme. The renormalization constant of the Fermi
  operator $Z_{O_F}$, although trivial ({\it i.e.} equal to one), has
  been included for generality.

\begin{figure}
    \parbox[t]{6.5cm}{\parbox[c][4cm][c]{6cm}{
	\psfig{file=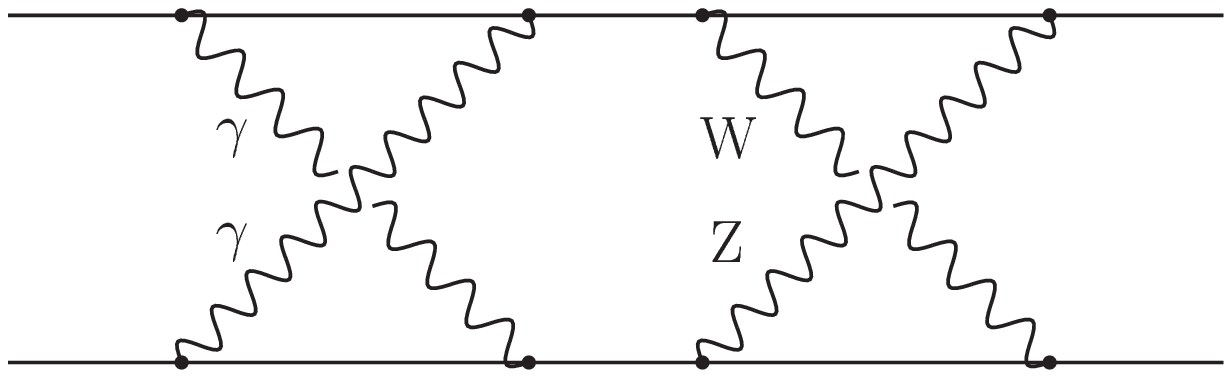,width=6.5cm}}
      \caption{\label{problem}
	A three loop diagram, which cannot be defined by Fierz symmetry
	with respect to the last line.}}
    \hspace{.5cm}
    \parbox[t]{6.5cm}{\parbox[c][4cm][c]{6cm}{
	\hspace{1.5cm}\psfig{file=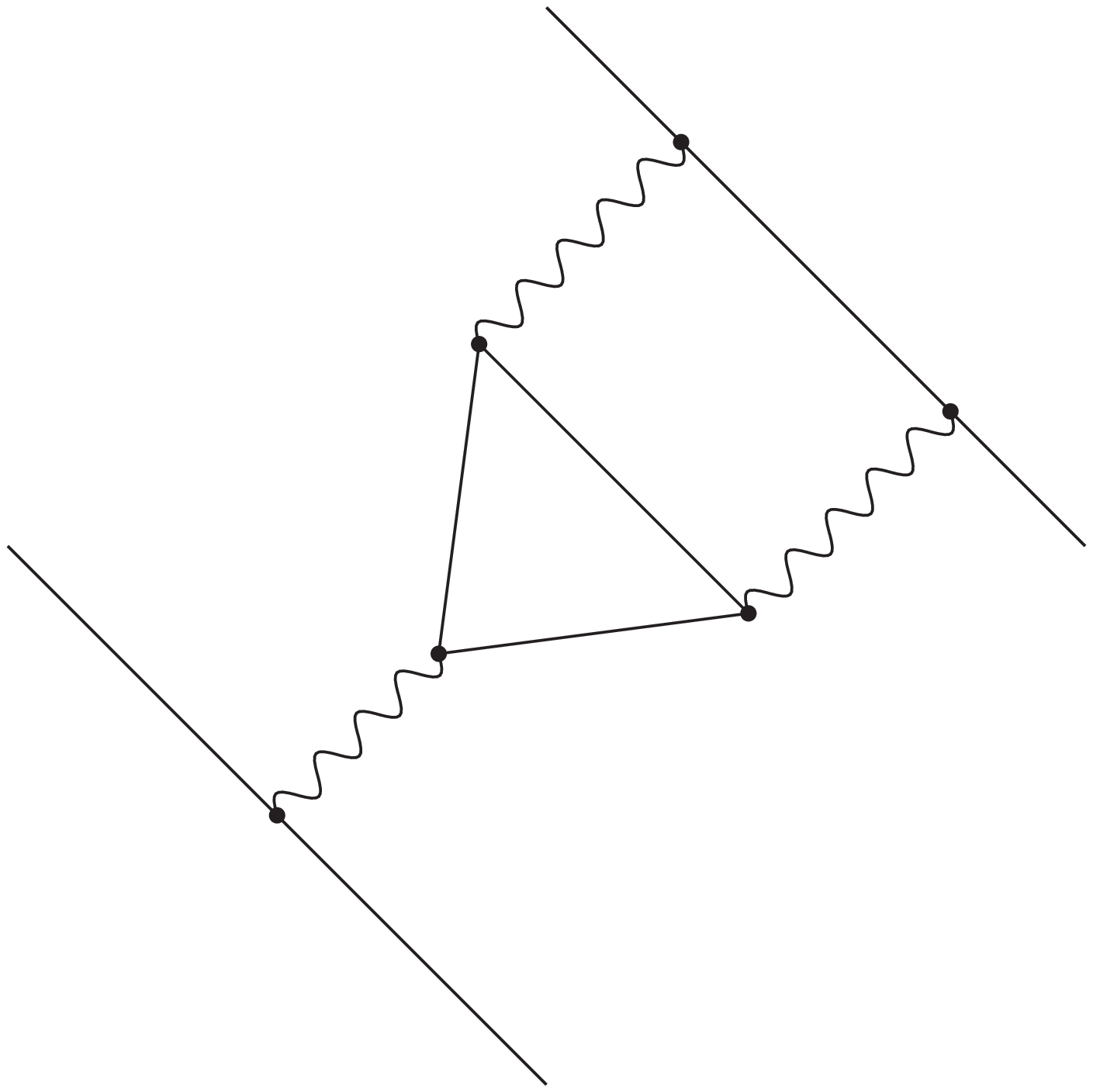,width=4cm}}
	\caption{\label{triangle}A triangular femion loop requiring
	  special treatment of the $\gamma_5$ matrix.}}
  \end{figure}

  The matching equation, Eq.~\ref{matching}, can be solved in
  different ways. The apparently simplest is to put all light masses
  and external momenta to zero, and renormalize the wave functions in
  the on-shell scheme. The right hand side in Fig.~\ref{matching1l}
  will then consist of only one term, proportional to $G_F^{(1)}$, if
  we use dimensional regularization, whereas the left will only have
  vacuum diagrams with heavy masses. Obviously, this situation will
  persist to all orders. The price to pay for this simplicity is the
  problem of infrared divergent box diagrams, where a product of gamma
  matrices occurs which does not have the form of the Fermi
  operator. In
  \cite{Awramik:2002wn,Onishchenko:2002ve,Awramik:2002vu}, this
  product has been defined through Fierz symmetry with respect to the
  last line in the string, which has been implemented, for practical
  reasons, by means of a suitable projection operator. Although
  sufficient at the two loop level, this symmetry will not suffice at
  the three loop level, see for example Fig.~\ref{problem}, where due
  to crossings, there is no last line in this sense.
  It is also not trivial that this procedure is correct even at the
  two loop level. For many topologies, {\it e.g.} those that contain a
  self energy insertion on the gauge boson line, one can convince
  oneself that this is indeed the case, others like the nonplanar
  double boxes in the purely bosonic contributions are not that
  easy. A highly nontrivial test of the calculation would consist in
  performing the matching without generating spurious infrared
  divergences\footnote{Note that one could also introduce evanescent
  operators \cite{Buras:1989xd} and perform the calculation with
  vanishing fermion masses and without projection, as for example in
  \cite{Misiak:1999yg}. This would be a second independent test.}. The
  box diagrams being ultraviolet finite can then be calculated in
  four dimensions, avoiding completely the problem of ambiguous gamma
  matrix definitions.

  In this work we performed the matching by keeping a common mass for
  all the light fermions and evaluating the box diagrams in four
  dimensions. It was necessary to calculate both sides of the matching
  equation and reexpand them subsequently in this common mass. The
  external wave function renormalization constants were not taken in
  the on-shell scheme, because this would introduce the usual on-shell
  infrared divergence. On the contrary, they were evaluated at zero
  momentum, which, in practice, is equivalent to renormalization in
  the $\overline{\rm MS}$ scheme with nonvanishing decoupling
  coefficients. Moreover, it turned out that it is necessary to have a
  correct $W$ boson wave function renormalization constant, since the
  box diagrams are not gauge invariant by themselves, and in the
  massive case this constant cancels only in combination with  vertex
  diagrams. We used a photon mass regulator, but the $\overline{\rm
  MS}$ renormalization constant would have been just as good. In the
  end, complete agreement was found with the calculation performed
  with massless fermions and with the projector conserving Fierz
  symmetry with respect to the last line from
  \cite{Awramik:2002wn,Onishchenko:2002ve,Awramik:2002vu}.

  \section{Results}

\begin{figure}
    \psfig{file=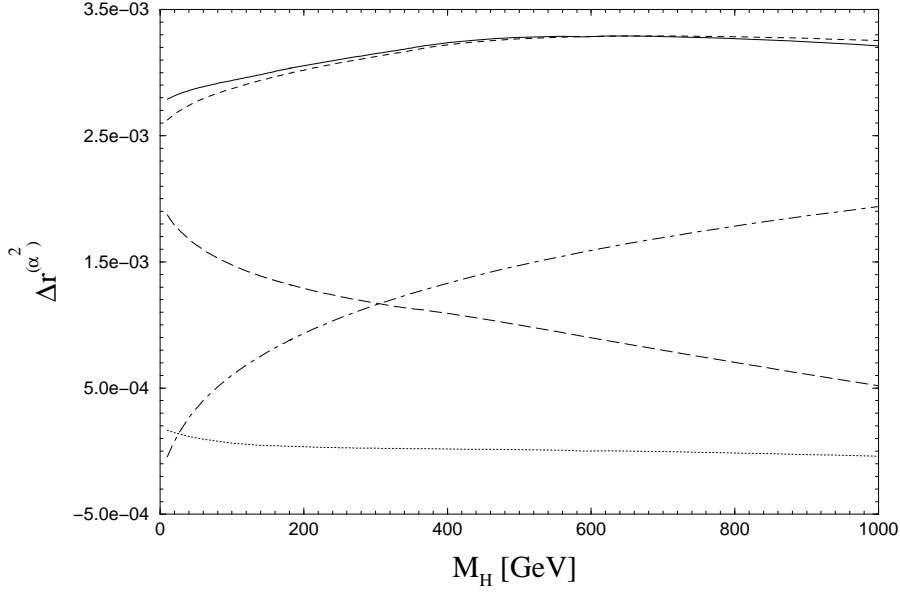,height=8cm,angle=270}
    \caption{\label{dr} Complete two loop electroweak contributions to
    $\Delta r$ (solid line) together with partial corrections: bosonic
    (dotted line), fermionic (dashed line), light fermionic without b
    quark, but with running of the fine sturcture constant
    (dash--dotted line) and top--bottom (long dashed line).}
  \end{figure}

  A detailed presentation of the methods used to evaluate the bosonic
  contributions to $\Delta r^{(\alpha^2)}$ can be found in
  \cite{Awramik:2002vu}. The fermionic contributions introduce two
  additional problems. First, some of the two loop vertex diagrams
  contain closed triangular fermion loops as shown in
  Fig~\ref{triangle}. The $\gamma_5$ matrix that occurs in the trace
  has to be correctly defined. We chose the naive dimensional
  regularization scheme \cite{Chanowitz:1979zu}, with an anticommuting
  $\gamma_5$ and the four  dimensional value of the trace of four
  gamma matrices and $\gamma_5$
  \begin{equation}
    {\rm Tr}(\gamma^\alpha \gamma^\beta \gamma^\gamma \gamma^\delta
    \gamma_5) = 4 i \epsilon^{\alpha\beta\gamma\delta}.
  \end{equation}
  This choice is justified by the fact that the nonvanishing
  contribution of the purely four dimensional $\epsilon$ tensors is
  finite. Moreover, it has been checked in \cite{Freitas:2000gg} that
  the use of the consistent definition of 't Hooft and Veltman
  \cite{'tHooft:fi} gives the same result after correction of the
  Green functions by suitable finite counterterms restoring the
  Slavnov--Taylor identities. Second, the inclusion of fermions
  results in unstable gauge bosons, which makes a proper definition of
  their masses necessary if gauge invariance of $G_F$ is to be
  maintained \cite{Sirlin:fd}. We use the pole mass scheme, where the
  inverse propagator matrix
  \begin{equation}
    (s-M_i^2)\delta_{ij}-\Pi_{T\;ij}(s),\;\;\;\; i,j=W,\gamma,Z,
  \end{equation}
  is singular in the complex $s$ plane at points which can be
  parametrized as
  \begin{equation}
    s_{\rm P} = M_{\rm P}^2-iM_{\rm P}\Gamma_{\rm P},
  \end{equation}
  where $M_{\rm P}$ is the mass and $\Gamma_{\rm P}$ is the width of
  the boson. This generates a fixed width Breit--Wigner behavior of
  the total cross section
  \begin{equation}
    \sigma (s) \sim \frac{1}{(s-M_{\rm P}^2)^2+M_{\rm P}^2 \Gamma_{\rm
    P}^2},
  \end{equation}
  as opposed to the running width parametrization actually used by the
  experimental collaborations for the masses and widths of the $W$ and
  $Z$ bosons \cite{Bardin:1988xt}
  \begin{equation}
     \sigma (s) \sim \frac{1}{(s-M_{\rm exp}^2)^2+s^2 \Gamma_{\rm
     exp}^2/M_{\rm exp}^2}.
  \end{equation}
  We translate back and forth between the two definition with the help
  of the following relations
  \begin{equation}
    \label{translation}
    M_{\rm P} = M_{\rm exp} \left( 1+\frac{\Gamma_{\rm exp}^2}{M_{\rm
    exp}^2} \right)^{-1/2},\;\;\;\; \Gamma_{\rm P} = \Gamma_{\rm exp}
    \left( 1+\frac{\Gamma_{\rm exp}^2}{M_{\rm exp}^2} \right)^{-1/2}.
  \end{equation}
  As in \cite{Freitas:2000gg}, we take $\Gamma_Z$ as experimentally
  measured, whereas we assume $\Gamma_W$ to be given by the one loop
  QCD corrected value
  \begin{equation}
    \Gamma_W = \frac{3 G_F M_W^3}{2 \sqrt{2} \pi} \left( 1 + \frac{2
    \alpha_s(M_W)}{3 \pi} \right).
  \end{equation}
  The complete result for $\Delta r$ at order $\alpha^2$ and the
  partial contributions are given in Fig.~\ref{dr}.
  The top quark mass and the running of the fine structure constant
  are taken from Tab.~\ref{input}, whereas the masses of the gauge
  bosons are translated from the experimental values given there to
  the pole mass scheme values with the help of Eq.~\ref{translation},
  which in this case gives $M_W = 80.424$~GeV and $M_Z = 91.1535$~GeV.
  \begin{table}
    \begin{center}
      \begin{tabular}{||l|l|l||}
	\hline
	\hline
	input parameter & value & source \\ 
	\hline 
	$M_W$ & $80.451 (33) {\rm \; GeV}$  & \cite{Hagiwara:fs} \\ 
	$M_Z$ & $91.1876 {\rm \; GeV}$ & \cite{Hagiwara:fs} \\ 
	$m_t$ & $174.3 (51) {\rm \; GeV}$ & \cite{Hagiwara:fs} \\
	$m_b$ & $4.7 {\rm \; GeV}$ & \cite{Freitas:2000gg} \\
	$G_\mu$ & $1.16637 \times 10^{-5} {\rm \; GeV}^{-2}$ & \cite{vanRitbergen:1998yd} \\
	$\alpha^{-1}$ & $137.03599976$ & \cite{Hagiwara:fs} \\
	$\Delta\alpha$ & $0.059228 (209) $ & \cite{Jegerlehner:2001wq}\\
	$\alpha_s(M_Z)$ & $0.119$ & \cite{Hagiwara:fs} \\
	$\Gamma_Z$ & $2.4952 {\rm \; GeV}$ & \cite{Hagiwara:fs} \\
	\hline
	\hline
      \end{tabular}
    \end{center}
    \caption{\label{input} Input parameters with experimental errors,
      where necessary for the present work. The value of $m_b$ is the
      same as in \cite{Freitas:2000gg} for comparison purposes.}
  \end{table}

  In order to compare our result for the fermionic contributions with
  \cite{Freitas:2000gg}, we evaluate the $W$ boson mass from the formula
  \begin{equation}
    M_W = M_Z \sqrt{\frac{1}{2}+\sqrt{\frac{1}{4}-\frac{\pi
    \alpha}{\sqrt{2} G_F M_Z^2}(1+\Delta r)}},
  \end{equation}
  with
  \begin{equation}
    \Delta r = \Delta r^{(\alpha)}+\Delta r^{(\alpha \alpha_s)}+\Delta
    r^{(\alpha \alpha_s^2)}+\Delta^{(\alpha^2)}_{\rm ferm}.
  \end{equation}
  We keep a finite $b$ quark mass in $\Delta r^{(\alpha)}$ and $\Delta
  r^{(\alpha \alpha_s)}$ and take the result for $\Delta r^{(\alpha
  \alpha_s^2)}$ from \cite{Chetyrkin:1995js}. Note also that we do not
  resum the running of the fine structure constant, {\it i.e.} $\Delta
  r^{(\alpha)}$ contains the term $+\Delta \alpha$ and $\Delta
  r^{(\alpha^2)}_{\rm ferm}$ includes $+\Delta \alpha^2$. The result
  is summarized in Tab.~\ref{comparison} for different Higgs boson
  masses from the range from 100~GeV to 1~TeV. We observe a
  discrepancy of around $-1.3$~MeV with respect to
  \cite{Freitas:2000gg}, which comes solely from the differing
  fermionic contributions\footnote{The authors of
  \cite{Freitas:2000gg} traced a problem in their calculation and
  after corrections agree with our results both for the fermionic and
  for the Higgs boson mass dependence of the bosonic two loop
  contributions. We checked that all of the remaining corrections are
  the same to required numerical accuracy.}.
  
  Inclusion of the bosonic corrections generates an additional
  variable shift already given in
  \cite{Awramik:2002wn,Onishchenko:2002ve}. As a result, our complete
  contributions induce a change of the $M_W$ prediction by $-2.4$~MeV
  for a Higgs boson mass as low as 100~GeV (see
  Tab.~\ref{comparison}). Since the bosonic part becomes negative for
  a heavier Higgs boson, this shift reaches $-0.6$~MeV for $M_H = 1$~TeV.
  \begin{table}
    \begin{center}
      \begin{tabular}{||c||lcc||cc||}
	\hline
	\hline
	& $\Delta r^{(\alpha)} +\Delta r^{(\alpha^2)}_{\rm ferm}$ 
	& $\; + \; \Delta r^{(\alpha \alpha_s)} + \Delta r^{(\alpha \alpha_s^2)}$ & 
	& $+ \Delta r^{(2)}_{\rm bos}$ & \\ 
	\hline
	\hline
	$M_H$ & $M_W$ \cite{Freitas:2000gg} & $M_W$ & $\Delta M_W$ & $M_W$ & $\Delta M_W$ \\
	$\rm [GeV]$ & [GeV] & [GeV] & [MeV] & [GeV] & [MeV] \\
	\hline
	100   & 80.3771  & 80.3758  &    -1.3  & 80.3747 & -2.4  \\
	200   & 80.3338  & 80.3326  &   -1.2  & 80.3321 & -1.7   \\
	600   & 80.2521  & 80.2509  &    -1.2  & 80.2508 & -1.3 \\
	1000  & 80.2135  & 80.2122  &  -1.3  & 80.2129 & -0.6 \\
	\hline
	\hline
      \end{tabular}
      \caption{\label{comparison} Comparison of the $M_W$ prediction
	(third and fifth column) against \cite{Freitas:2000gg} (second
	column). $\Delta M_W$ is, in both cases, the shift with
	respect to the fitting formula.}
    \end{center}
  \end{table}

\begin{figure}
    \psfig{file=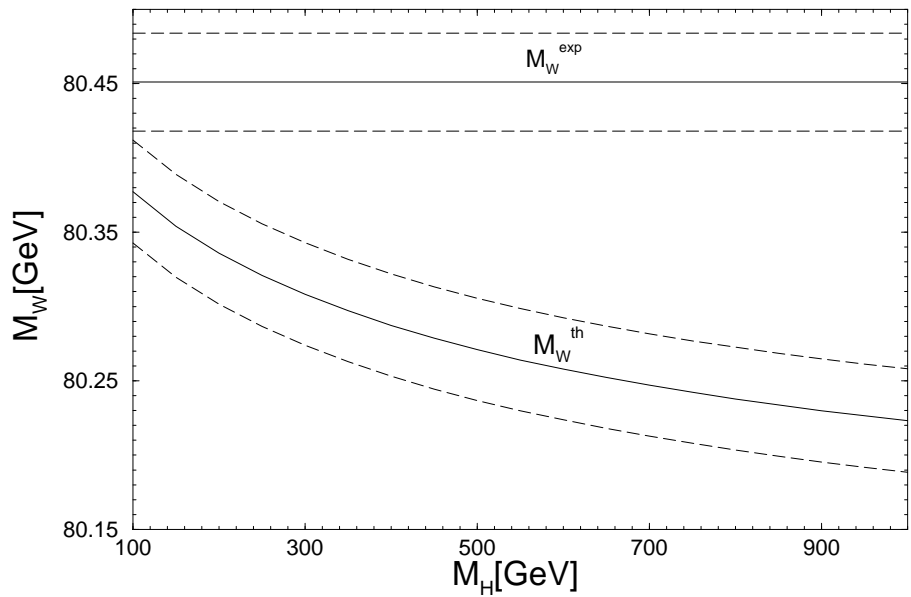,height=8cm,angle=270}
    \caption{\label{mw} The theoretical prediction for the $W$ boson
    mass, $M_W^{\rm th}$, with error from the uncertainty of the top
    quark mass and the running of the fine structure constant, against
    the current experimental value, $M_W^{\rm exp}$.}
  \end{figure}

  In Tab.~\ref{full}, we include also the recent partial results at
  three loop order \cite{Faisst:2003px}, {\it i.e.} we use
  \begin{equation}
    \Delta r = \Delta r^{(\alpha)}+\Delta r^{(\alpha \alpha_s)}+\Delta
    r^{(\alpha \alpha_s^2)} + \Delta^{(\alpha^2)} -
    \frac{c_W^2}{s_W^2}\left(\Delta \rho_t^{(\alpha^3)} + \Delta
    \rho_t^{(\alpha^2 \alpha_s)}\right).
  \end{equation}
  \begin{table}
    \begin{center}
      \begin{tabular}{||c||l|cc||cc||}
	\hline
	\hline
	& & $ - c_W^2/s_W^2 \Delta \rho^{(\alpha^3)}_{\rm t}$
	& & $ - c_W^2/s_W^2  \Delta \rho^{(\alpha^2\alpha_s)}_{\rm t}$ & \\ 
	\hline
	\hline
	$M_H$ & $M_W$ & $M_W$ & $\Delta M_W$ & $M_W$ & $\Delta M_W$ \\
	$\rm [GeV]$ & [GeV] & [GeV] & [MeV] & [GeV] & [MeV] \\
	\hline
	100 & 80.3747 & 80.375 & 0.3 & 80.3771 & 2.4 \\ 
	200 & 80.3321 & 80.3322 & 0.1 & 80.3358 & 3.7 \\ 
	600 & 80.2508 & 80.2510 & 0.2 & 80.2579 & 7.1 \\ 
	1000 & 80.2129 & 80.2146 & 1.7 & 80.2231 & 10.2 \\
	\hline
	\hline
      \end{tabular}
      \caption{\label{full} Additional shift of $M_W$ with respect to
	the complete prediction from Tab.~\ref{comparison} due to
	inclusion of partial results at order $\alpha^3$ and $\alpha^2
	\alpha_s$ from \cite{Faisst:2003px}.}
    \end{center}
  \end{table}
  Together with errors coming from the top quark mass and the running
  of the fine structure constant but without a theoretical error
  estimate, the $M_W$ prediction is shown against the current
  experimental result in Fig.~\ref{mw}.

  \section{Conclusions}

  We have presented a new result for the complete electroweak
  contributions to the lifetime of the muon, which induces a shift in
  the $W$ boson mass prediction as large as $-2.4$~MeV for a light
  Higgs boson, of which $-1.3$ MeV come from a discrepancy with the
  previous calculation of the fermionic contributions
  \cite{Freitas:2000gg} and the rest from the bosonic part. The
  authors of \cite{Freitas:2000gg} corrected their
  evaluation\footnote{see the updated version \cite{update}} and are
  now in full agreement with this work. Together with recent results
  at the three loop level \cite{Faisst:2003px}, this calls for an
  updated fitting formula. Such a formula will be given in a
  subsequent publication \cite{fit}.

  \begin{ack}
    
    The authors would like to thank A. Freitas for his effort put into
    the comparison of the present results and the results of
    \cite{Freitas:2000gg} and for tracing an error in an earlier
    version of this paper, and G. Weiglein for reading the
    manuscript. The warm hospitality of the Institute for Particle
    Physics Phenomenology of the University of Durham  during the time
    when part of this work was completed is gratefully
    acknowledged. This work was supported in part by the KBN Grant
    5P03B09320.

  \end{ack}

\end{document}